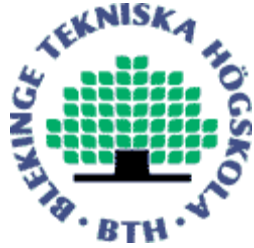

# What do we know about software development in startups?

Carmine Giardino, Michael Unterkalmsteiner, Nicolò Paternoster, Tony Gorschek, Pekka Abrahamsson

Startups are newly created companies with little or no operating history facing a high volatility in technologies and markets. In the US alone, 476,000 new businesses are established each month [1], accounting for about 20% of job creation [2]. As such, startups are an important factor in the economy. However, the environment of startups is dynamic, unpredictable and even chaotic, forcing entrepreneurs to act quickly, fail fast and learn faster to find a market niche to acquire a sustainable income. 60% of the startups do not survive the first five years, and 75% of venture capital funded startups fail [3]. Most of this is due to the high risk of startups, missed market window, and other business reasons. However, since engineering practices of startups are largely unknown, to what extent inadequate engineering is a contributing factor to this high failure rate is also largely speculation.

We present a detailed investigation and collection of all known empirical software engineering sources related to startups and their engineering practices, as well as analysis of how accurate reliable this available evidence is [4]. We see this as a first critical step into an area largely unknown, namely the world of software engineering practices in startups.

## What is a startup, anyway?

In the past, the term "startup" has been associated with different meanings. Looking at the recurrent themes (see Table 1 for a complete list) adopted by researchers and practitioners, a startup is a small company exploring new business opportunities, working to solve a problem where the solution is not well-known and the market is highly volatile. Being newly founded does not in itself make a company a startup. High-uncertainty and rapidly-evolvement are the two key characteristics for startups retrieved by the studies, which better differentiate them from more established companies.

*Table 1: Recurrent themes in software startups*

| Theme | Description |
|---|---|
| Lack of resources | Economical, human, and physical resources are extremely limited. |
| Highly reactive | Startups are able to quickly react to changes of the underlying market, technologies, and product (compared to more established companies) |
| Innovation | Given the highly competitive ecosystem, startups need to focus and explore highly innovative segments of the market. |
| Uncertainty | Startups deal with a highly uncertain ecosystem under different perspectives: market, product features, competition, people and finance. |
| Rapidly evolving | Successful startups aim to grow and scale rapidly. |

| Time-pressure | The environment often forces startups to release fast and to work under constant pressure (terms sheets, demo days, investors' requests). |
|---|---|
| Third party dependency | Due to lack of resources, to build their product, startups heavily rely on external solutions: External APIs, Open Source Software, outsourcing, COTS, etc. |
| Small team | Startups start with a small numbers of individuals. |
| One product | Company's activities gravitate around one product/service only. |
| Low-experienced team | A good part of the development team is formed by people with less than 5 years of experience and often recently graduated students. |
| New company | The company has been recently created. |
| Full organization | Startups are usually founders-centric and everyone in the company has big responsibilities, with no need of high-management. |
| Highly risky | The failure rate of startups is extremely high. |
| Not self-sustained | Especially in the early stage, startups need external funding to sustain their activities (Venture Capitalist, Angel Investments, Personal Funds, etc.). |
| Little working experience | The basis of an organizational culture is not present initially. |

# Empirical body of evidence (SIDEBAR)

A Systematic Mapping Study is a method to structure the empirical evidence in a particular field of interest [5]. We identified 43 studies that investigate different aspects of startups and their software development processes. We also estimated the strength of evidence in this field by assessing the rigor and relevance [6] of the studies (Figure 1). Rigor refers to the precision and thoroughness of reporting the design, validity threats and results of a study. Relevance refers to the realism of the environment where the study is performed and to the potential of transferring the results to practitioners.

The rigor and relevance assessment suggests that the empirical evidence on the startup phenomenon is still rather premature. A minority, 10 of the 43 mapped studies, provides transferable and reliable results to practitioners (sector A). Similarly, 10 studies provide low rigor and relevance (sector C). More studies (23) exhibit moderate industry relevance, however with low scientific rigor (sector B). From this observation, we conclude that it is challenging to conduct research in an environment where lack of resources is one of the dominant characteristics. Researchers need to identify efficient means to collaborate with and study startups.

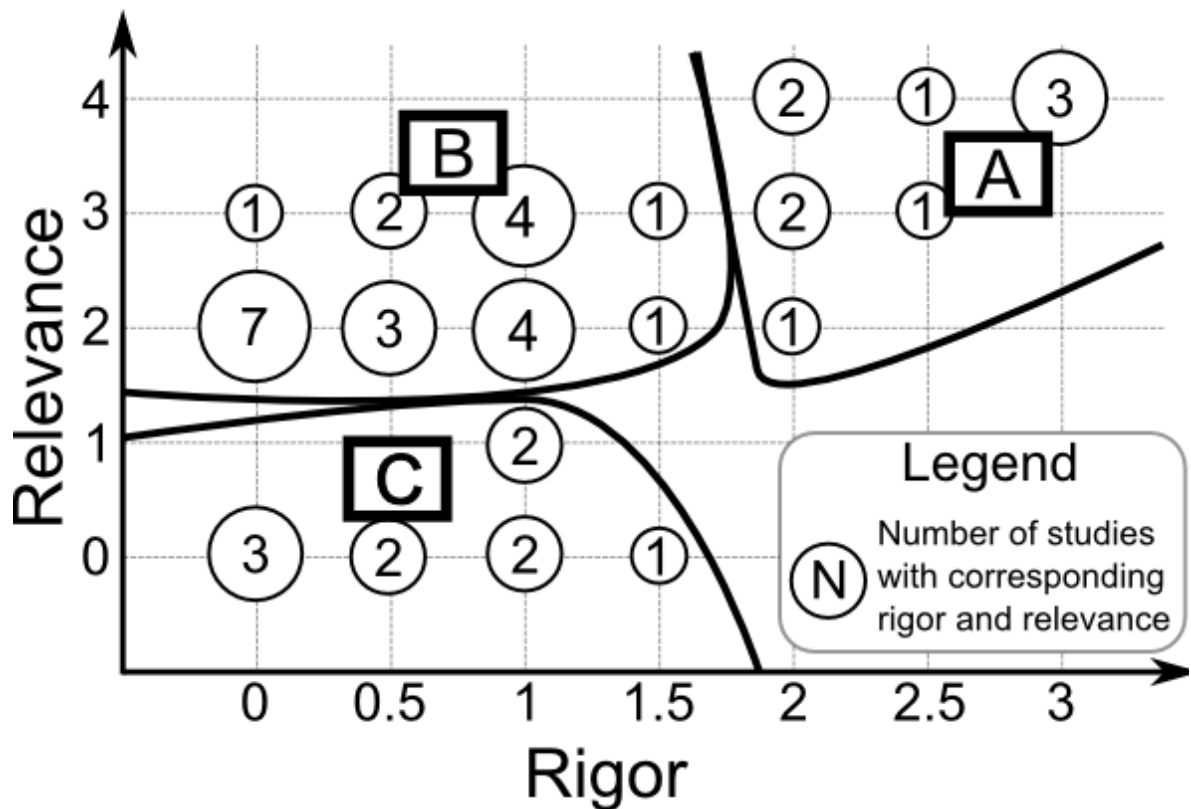

*Figure 1: Rigor and relevance of primary studies*

# Startup software development

"Done is better than perfect" or "move fast and break things" are slogans you can read when entering in a startup working space. What stands behind those slogans is a summary of more than 200 working practices. We summarized evidences and pointed out where gaps exist and future development and research are warranted.

## Process management is agile, evolutionary and opportunistic

Process management represents all the engineering activities used to manage product development in startups. As the flexibility to accommodate frequent changes is essential in the startup context, agile methodologies have been considered as the most viable process since they embrace change, allowing development to adapt to the business strategy [7]. Fast release with an iterative and incremental approach shortens the lead time from idea conception to production with fast deployment.

A variant to agile is the Lean Startup methodology [8], which advocates the identification of the most risky parts of a software business and provide a minimum viable product to systematically test and plan modification for the next iteration. In this regard, in order to shorten time-to-market, prototyping is essential.

In order to allow better prototyping activities, evolutionary workflows are needed to implement "soft-coded" solutions in the first phases until the optimal solution is found. Despite the amount of methodologies which embrace fast prototyping in the development, evidence shows that none of the processes are strictly followed by startups. Yet, the uncertainty and fast changing needs in startups drive them to opportunistically tailor minimal process management to their short-term objectives and adapt to the fast-paced learning process of their users to address the uncertainty of the market.

## Software development is driven by customers, who act as designers

Startups are under constant pressure to demonstrate rapidly that they are developing a solution that fixes a real problem [9]. We can say that startuppers are in a constant process of optimizing the problem/solution fit. In order to achieve this, startups discover the real needs of their first customers, testing business speculations only by defining a minimal set of functional requirements [10].

Several authors acknowledge the importance of involving the customer/user in the process of eliciting and prioritizing requirements according to their primary needs. However the market-driven nature of the requirements demands also for alternatives. For example, startups can use scenarios in order to identify requirements in the form of user stories and estimate the effort for each story. However, polishing requirements that address an unsolicited need are a waste. Requirements elicitation methods are moving towards testing the problem and understanding if the solution fits to real needs even before the product goes to the market (e.g. the customer development process [9]).

In the startup context, customers often steer the requirements and the developers need to be ready to embrace the change from day one. The use of architecture and design patterns to make features modular and independent is seen crucial when functionality is continuously been updated or removed. Therefore, employing architectural practices and frameworks that enable easy extension of the design can dramatically benefit the alignment between the product and the uncertainty of the market needs [11]. This requires some upfront effort but can prevent the growth of product complexity before obtaining initial revenue.

Scientific evidence also points out to advantages of constant code refactoring. Need of re-implementation of the whole system might be costly and risky if it needs to be urgently scalable to a growing number of users. Therefore, some quality assurance is needed for the functionality that brings the most value to the customers. The use of ongoing customer acceptance with the use of focus groups of early adopters can provide a time-efficient mean to discover major bugs. Solutions are still however scarce for finding easily accessible automated testing frameworks and also for more practical user interface testing approaches.

## Team is the catalyst of development

Time pressure and lack of resources lead startups to a loose organizational structure and often lack of traditional management [12]. Empowerment of team members represents the main viable strategy to enhance performance and chances of success [13]. The team needs to be able to absorb and learn from trial and error quickly enough to adapt to new emergent practices. Working on innovative products requires creativity, ability to adapt to new roles and to face new challenges everyday, working overtime if necessary.

Indeed, in building up a startup company, the team needs expertise to counterbalance the lack of resources. Previous experience in similar business domains and a working history in a team, exhibiting characteristics of an entrepreneur (courage, enthusiasm, commitment, leadership), also play a primary role in the skill set of a startup employee. Nevertheless, the absence of structure might hinder important activities, such as sharing knowledge and team coordination, especially when the company grows. In this case co-location is essential to facilitate informal communication and close interactions between team-members.

**Tools are ready to accommodate product and management changes**

Startups are in the position where they can take advantage of using newest technologies and development tools without having any legacy or being constrained by previous working experience [14]. The selection of a technology requires some domain-specific or product-specific requirements, which are however mostly unknown in the early-stages, posing an additional challenge for technology selection.

In general, startuppers prefer those technologies that can quickly accommodate change in the product and its management [15]. Examples include general-purpose infrastructures, such as configuration management, problem reporting, tracking and planning systems, scheduling and notification systems. Easy-to-implement tools, such as whiteboards and real-time tools to handle fast-paced changing information, lower training and maintenance costs. In order to mitigate the lack of resources, startups often appear to take advantage of open source solutions when possible, which also give access to a large pool of evaluators and evolving contributions.

# Practical takeaways

We studied the empirical evidence in software development in startups' context showing how software startups operate with high level of uncertainty. To quickly validate the product in the market, startups tend to use agile and lean methods in an ad-hoc manner [16]. Startup companies are seeking to generate revenue and obtain funding to continue the development, which means in practice that the software quality is not the most critical concern.

Evidence suggests that engineering activities need to be tailored to the startup context in order to allow flexibility and reactiveness in development workflows. Decision makers in startups confront with the continuous unpredictability of their startup endeavor. The relationship between cause and effect can only be perceived in retrospect [17]. Applying rigorous methodologies to control development activities is not effective, because no matter how much time is spent in analysis, it is not possible to identify all the risks or accurately predict what practices are required to develop a product.

On the other hand, flexible and reactive methods, designed to stimulate customer feedback, increase the number of perspectives and solutions available to decision makers. Acting fast and opportunistically, according to the context of development, justifies the need of discovering what a paying customer value proposition is and opens up new possibilities for creation, generating the condition for sustainable innovation. In other words, developers should have the freedom to choose activities quickly, stop immediately when the results are wrong, fix the approach and learn from the previous failures. In line with the Lean Startup movement, we expect methodologies and techniques tailored from common Agile practices to specific startups' culture and needs where failures are completely acceptable or even preferred in favor of a faster learning process.

Reported practices, which ride the wave of rapidly-evolving technologies and markets, are:

- Use of well-known frameworks providing fast changeability of the product according to market needs.

- Use of evolutionary prototyping and experimentations by means of existing components.

- Ongoing customer acceptance by early adopters' focus groups.

- Continuous value delivery, focusing on core functionalities which engage paying customers.

- Empowerment of team responsibilities and their ability to influence the final outcomes.

- Use of metrics to fast learn from consumers' feedback and demand.

- Use of easy-to-implement tools facilitating product development and its management, handling fast-paced changing information.

In conclusion, the startups of today are in the forefront of applying new technologies in practice. We see that the increasing startup phenomenon opens uncharted opportunities but also challenges in research. We place a call for more transferable and reliable results concerning the diversity of context and viewpoints in the adoption of practices dealing with high uncertainty.